\documentclass[twocolumn,jgrga]{agutex}
\usepackage{url}                                                

\usepackage[dvips]{graphicx}
\authorrunninghead{JANARDHAN ET AL.}

\titlerunninghead{Photospheric Fields: A 20 year Decline}

\authoraddr{Corresponding author:P. Janardhan, Astronomy \& Astrophysics Division, 
Physical Research Laboratory, Ahmedabad 380009, India. 
(e-mail: jerry@prl.res.in)}
\def\nat{{\it Nature,}}
\def\jgr{{\it Journal of Geophysical Research (Space Physics)}}
\def\grl{{\it Geophys. Res. Lett.,}}
\def\mnras{{\it Mon. Not. R. Astron. Soc.}}
\def\apj{{\it Astrophys. J.}}
\def\aa{{\it Astron. \& Astrophys. Lett.}}
\def\aap{{\it Astron. \& Astrophys.}}

\def\ssr{{\it Space Sci. Rev.}}
\def\solphys{{\it Sol. Phys.}}

\begin{document}

%
%

\title{{\bf{A Twenty Year Decline in Solar Photospheric Magnetic Fields: Inner-Heliospheric  
Signatures and Possible Implications?}}}

%
%


\authors{P. Janardhan,\altaffilmark{1}
Susanta Kumar Bisoi,\altaffilmark{2} 
S. Ananthakrishnan,\altaffilmark{3} 
M. Tokumaru,\altaffilmark{4} 
K. Fujiki,\altaffilmark{4} 
L. Jose,\altaffilmark{5} and 
R. Sridharan,\altaffilmark{5}}

\altaffiltext{1}{Astronomy \& Astrophysics Division, Physical Research 
Laboratory, Ahmedabad 380009, India.}
\altaffiltext{2}{National Astronomical Observatories, 
Chinese Academy of Sciences, Beijing 100012, China.}
\altaffiltext{3}{Department of Electronic Science, University of Pune, 
Pune - 411 007, India.}
\altaffiltext{4}{Solar-Terrestrial Environment Laboratory, Nagoya 
University, Nagoya 464-8601, Japan.}
\altaffiltext{5}{Space \& Atmospheric Sciences Division, Physical 
Research Laboratory, Ahmedabad 380009, India.}
%
%


\begin{abstract}
We report observations of a steady 20 year decline of solar photospheric fields at 
latitudes $\geq$45${^{o}}$ starting from $\sim$1995.  This prolonged and continuing 
decline, combined with the fact that Cycle 24 is already past its peak, implies that 
magnetic fields are likely to continue to decline until $\sim$2020, the expected 
minimum of the ongoing solar Cycle 24.  In addition, interplanetary scintillation (IPS) 
observations of the inner heliosphere for the period 1983--2013 and in the distance 
range 0.2--0.8 AU, have also shown a similar and steady decline in solar wind 
micro-turbulence levels, in sync with the declining photospheric fields.  Using 
the correlation between the polar field and heliospheric magnetic field (HMF) at 
solar minimum, we have estimated the value of the HMF in 2020 to be 3.9 ($\pm$0.6) 
and a floor value of the HMF of $\sim$3.2 ($\pm$0.4) nT.  Given this floor value 
for the HMF, our analysis suggests that the estimated peak sunspot number for solar 
Cycle 25 is likely to be $\sim$ 62 ($\pm$12).
\end{abstract}

%
%

%

\begin{article}

%
%
\section{Introduction}
     \label{S-Intro}
Sunspots or dark regions of strong magnetic fields on the Sun are generated 
via magneto-hydrodynamic processes involving the cyclic generation of toroidal, 
sunspot fields from pre-existing poloidal fields and their eventual regeneration 
through a process, referred to as the solar dynamo. This leads to the well 
known periodic 11-year solar cycle with sunspot numbers climbing, at solar maximum, 
up to 200 in very active or strong solar cycles and dropping down, at solar 
minimum, to 20 or less sunspots during periods of solar minimum.   

The current solar Cycle 24, on the one hand, was preceded by one of the deepest 
solar minima experienced in the past 100 years causing Cycle 24 to not only 
start $\sim$1.3 years later than expected \citep{JRL11}, but also be the weakest 
since solar Cycle 14 in the early 1900's.  With a peak smoothed monthly
sunspot number $\sim$75 in November 2013, the maximum of solar Cycle 24 has in fact 
been dubbed the ``mini" solar maximum \citep{McA13}.  On the other hand, 
measurements of the sunspot umbral field strength have been shown to be steadily 
decreasing by $\sim$50 G per year since $\sim$1998 \citep{PLi06,LPS12}.  Also the 
sunspot formation fraction, the ratio of the observed monthly smoothed sunspot 
number (SSN) and the SSN value derived from the monthly F10.7cm radio flux, have 
been reported to be steadily decreasing since $\sim$1995 \citep{LPS12}.  It is 
important to bear in mind that for field strengths below about 1500 G, there will 
be no contrast between the photosphere and sunspot regions \citep{LPS12}, thereby 
making sunspots invisible.  These authors have claimed that the umbral field strengths 
in Cycle 25 would be around 1500 G, and thus, there would be very little/no sunspots 
visible on the solar photosphere in the next solar Cycle 25. Studies of the heliospheric 
magnetic field (HMF), using ${\sl{in-situ}}$ measurements at 1 AU, have also shown a 
significant decline in strength \citep{SBa08,WRS09,CSS11,CLi11}.  It has been reported 
that during the so called ``mini" maximum of Cycle 24, the HMF intensity was more 
like that during the minimum of Cycle 23 recording an all-time-low space age value. 

%
\begin{figure*}
\centering
\includegraphics[height=5.0cm,width=10cm]{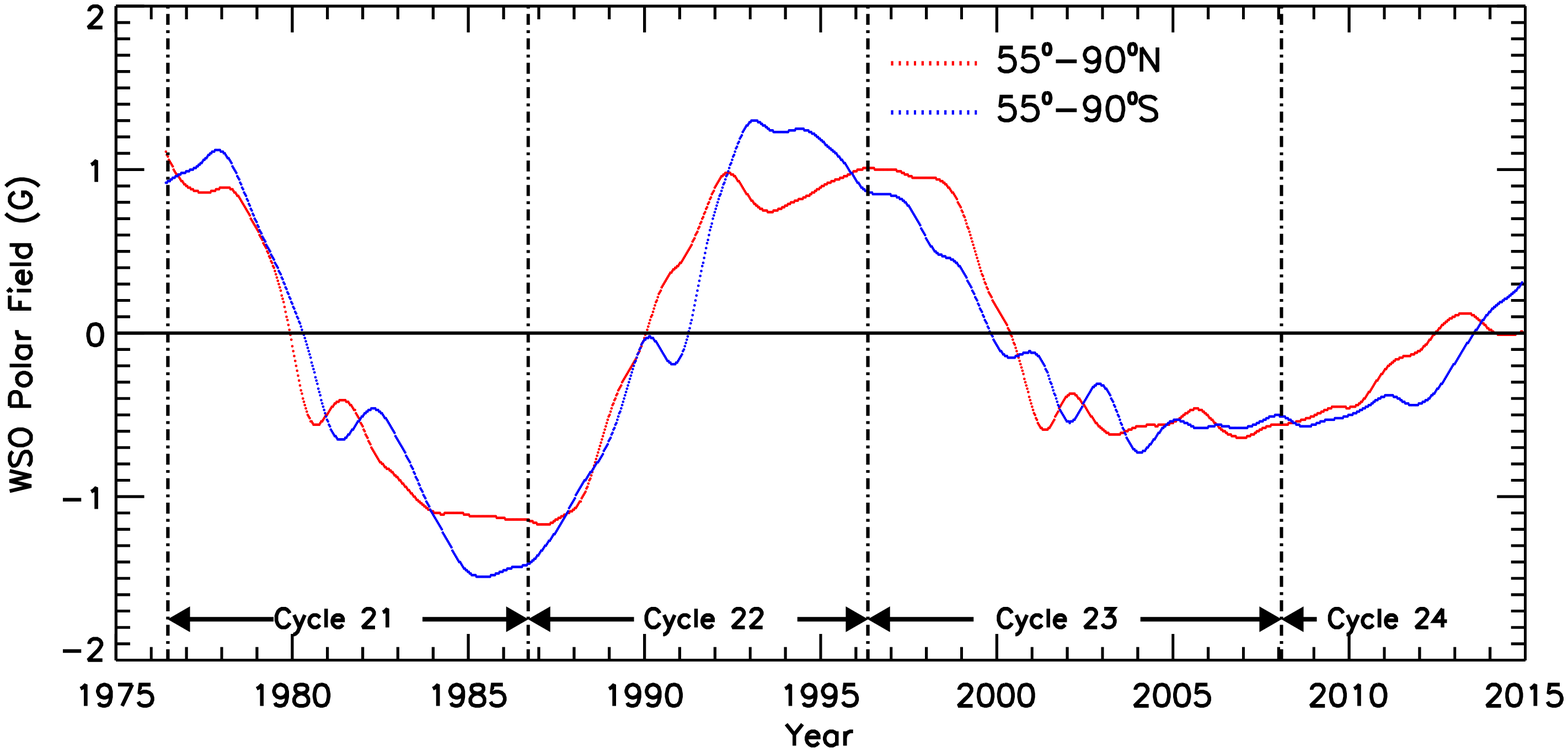}
\caption{The polar cap fields derived from the Wilcox Solar Observatory (WSO) data 
for the Northern (red) and Southern (blue) hemispheres for the period 1976.42\,--\,2014.91, 
covering solar cycles 21, 22, 23 and 24.  The Southern hemisphere fields have been inverted 
for ease of comparison.}
\label{fig1}
\end{figure*}
%

Recent studies, using synoptic magnetograms from the National Solar Observatory (NSO), 
Kitt-Peak (NSO/KP), USA, have reported a steady decline in photospheric magnetic fields 
at helio-latitudes ${\geq{45^{\circ}}}$, with the observed decline having begun in the 
mid-1990's \citep{JBG10}. Using 327 MHz observations from the four station IPS observatory 
of the Solar-Terrestrial Environment Laboratory (STEL), Nagoya University, Japan, we have 
examined solar wind micro-turbulence levels in the inner-heliosphere and have found a 
similar steady decline, continuing for the past 18 years, and in sync with the declining 
photospheric fields \citep{JaB11}. Our present work has confirmed that this declining trend 
is continuing to the present.  A study, covering solar Cycle 23, of the solar wind density 
modulation index, $\epsilon_{N} \equiv \Delta{N}/N$, where, $\Delta{N}$ is the rms electron 
density fluctuations in the solar wind and N is the density, has reported a decline of around 
8\% in the distance range 0.2 AU to 0.8AU \citep{BiJ14b}.  This decline has been attributed 
by the authors to the steadily declining solar photospheric fields. In light of the very 
unusual nature of the minimum of solar Cycle 23 and the current very weak solar Cycle 24, 
we have re-examined in this paper, both solar photospheric magnetic fields and the HMF 
for the period 1975\,--\,2014 and solar wind micro-turbulence levels during 1983\,--\,2013.  

The aim of this paper is to try and estimate the maximum strength of solar Cycle 25 and 
address the question of whether we are headed towards a long period of little/no sunspot 
activity similar to the well known Maunder minimum (1645-1715 AD) when the sunspot activity 
was extremely low.

\section{Data}
\subsection{Magnetic Field}
Magnetic field measurements from ground based magnetograms are freely available in 
the public domain and we have made use of these observations in this paper for 
estimating solar magnetic fields.  We have used photospheric magnetic fields 
from the Wilcox Solar Observatory (WSO), the National Solar Observatory at Kitt Peak, 
USA (NSO/KP), and the Synoptic Long-term Investigation of the Sun (SOLIS) facility at 
Kitt Peak, USA (NSO/SOLIS).  For estimating polar fields at latitudes from 
55${^{\circ}}$\,--\,90${^{\circ}}$,  we have used data obtained from 
the Wilcox Solar Observatory (WSO)(\url{http://wso.stanford.edu/Polar.html}), 
covering the period from 31 May 1976 (1976.42)\,--\,28 Nov 2014 (2014.91).  The polar 
field strength at WSO are line-of-sight magnetic fields measured each day in the 
polemost apertures for the Northern and Southern solar hemispheres calculated for every 
10 days.  A 20 nHz low pass filter is used to remove the yearly projection 
effects from the measured line-of-sight magnetic fields.   For the estimation of photospheric 
fields in the latitude range 0$^{\circ}$\,--\,45$^{\circ}$ and 45$^{\circ}$\,--\,78$^{\circ}$, 
synoptic magnetograms were used from NSO/KP ({\url{ftp://nsokp.nso.edu/kpvt/synoptic/mag/}}) and 
NSO/SOLIS ({\url{ftp://solis.nso.edu/synoptic/level3/vsm/merged/carr-rot/}}), respectively. 
The data are available as standard FITS files in the longitude and latitude format of 
360 $\times$ 180 arrays, for the period February 1975 (1975.14)\,--\,July 2014 (2014.42), 
covering Carrington Rotations (CR) CR1625\,--\,CR2151. 

For the study of heliospheric magnetic fields (HMF), we have used 27-day averaged HMF 
data at 1 AU, available in the public domain from the OMNI2 database, for the period 
1975\,--\,2014. The OMNI data base is a compilation of hourly-averaged, near-Earth 
solar wind magnetic field and plasma parameter data from several spacecraft in geocentric 
or L1 (Lagrange point) orbits.
\subsection{IPS measurements}
For the study of solar wind micro-turbulence levels in the inner heliosphere, 
we have used the daily IPS measurements of scintillation indices, spanning the 
period from 1983 to 2013. The IPS observations on a set of about 200 chosen 
extra-galactic radio sources have been carried out on a regular basis at the 
multi-station IPS observatory of STEL, Japan to determine scintillation indices 
at 327 MHz \citep{KKa90,AsK98} since 1983 till date.  Prior to 1994, however, 
these observations were carried out only by the three-station IPS facility at 
Toyokawa, Fuji, and Sugadaira.  One more antenna was commissioned at Kiso in 
1994, forming a four\,--\,station dedicated IPS network, that has been making 
systematic and reliable estimates scintillation indices \citep{TKF12}.  Each day 
about a dozen selected radio sources have been observed such that each 
source would have been observed over the whole range of heliocentric distances 
between 0.2 and 0.8 AU in a period of about 1 year.  

\section{Photospheric Magnetic Fields}
\label{S-mag-field}

While studying solar polar fields, different groups of researchers have considered 
different solar latitude ranges to represent solar high latitude fields or polar fields, 
by averaging the magnetic flux in the defined high latitude polar regions. Till date 
the polar area considered has been arbitrary \citep{UHa14}.  For example, Wilcox Solar 
Observatory (WSO) polar fields are those at latitudes $\geq$55$^{\circ}$.  Other researchers 
have used the area above latitudes $\geq$60$^{\circ}$ \citep{Tom11} and still others have 
used the area above latitudes $\geq$70$^{\circ}$ \citep{MuS12} to represent polar or 
high latitude fields.  
\begin{figure*}
\centering
\includegraphics[height=10.0cm,width=10cm]{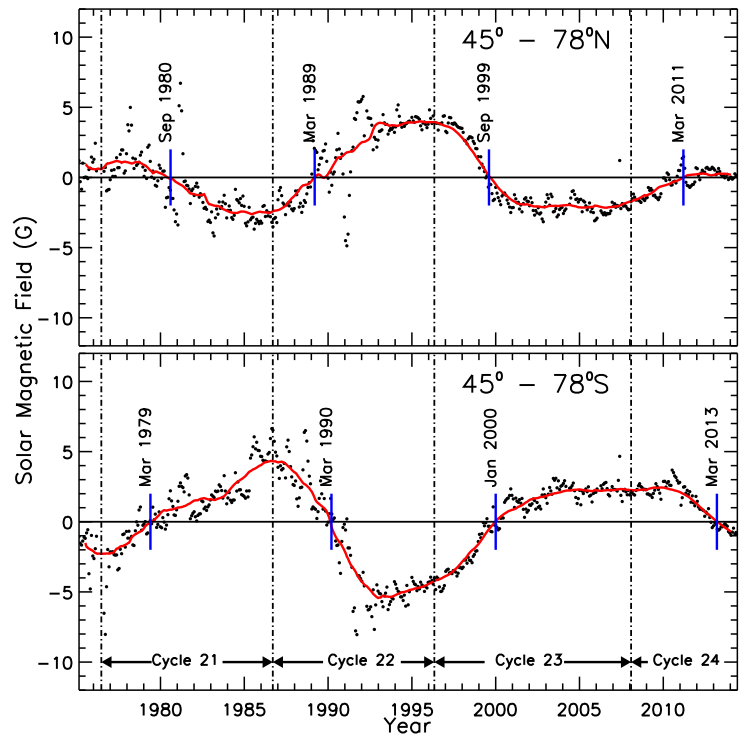}
\caption{The signed photospheric magnetic fields in the latitude 
range 45${^{\circ}}$ to 78${^{\circ}}$ for the Northern (top) and Southern 
(bottom) hemispheres for solar cycles 21, 22, 23 and 24.  The solid black 
dots are actual magnetic measurements from the NSO/KP magnetograms, while 
the solid red curve depicts the smoothed magnetic fields.  The time of 
reversal of magnetic fields for each Solar Cycle in both the hemispheres 
is indicated by a small blue arrow.}
\label{fig2}
\end{figure*}

For the present study, we have considered photospheric magnetic fields in the latitude 
range 45$^{\circ}$\,--\,78$^{\circ}$ as representative of polar fields, referred to 
in the rest of the paper as high-latitude fields.  Using NSO/KP data, available in 
the public domain as described earlier, photospheric magnetic fields were estimated by 
a longitudinal average of the whole 360${^\circ}$ array of longitude to produce 1${^\circ}$ 
strips of data for each CR of 27.275 days.  High-latitude fields were then computed 
by appropriate averaging \citep{JBG10,BiJ14}.

It is known that the (signed) polar field or polar flux normally reverses polarity at 
each solar maximum, commonly referred to as ``polar field reversals or polar reversals" 
\citep{Bab59,Bab61}.  For example, if the polar field in the Northern (Southern) 
hemisphere has positive (negative) polarity, then after polar reversal it will have 
negative (positive) polarity.  The solar polar fields reverse because the excess amount 
of trailing polarity flux from decaying sunspots move polewards, cancel the polar cap 
fields of opposite polarity, and impart a new polarity.  It has been seen 
however, that the reversal is hemispherically asymmetric or in other words, occurs at 
different times in the Northern and Southern hemispheres \citep{SKa13}.  Figure \ref{fig1} 
shows the signed polar cap fields derived from WSO data for the Northern (red) and 
Southern (blue) solar hemisphere with the Southern polar cap fields being inverted for 
ease of comparison.  The hemispherical asymmetry in the polar field reversals in each
solar hemisphere for solar cycles 21, 22, 23 and 24, in the latitude range 
45$^{\circ}$\,--\,78$^{\circ}$, is clearly seen from Fig. \ref{fig1}.

The upper and lower panel of Figure \ref{fig2} shows the signed value of the 
high-latitude fields for the solar Northern and Southern hemispheres respectively, 
for solar cycles 21, 22, 23, and 24.  The solid black dots are actual magnetic 
measurements derived using NSO/KP magnetograms, while the solid red curve shows 
the variation of the smoothed magnetic fields.  A field reversal is deemed 
to have taken place when the solid red curve in Fig. \ref{fig2} changes sign. 
The small blue lines in both panels of Fig. \ref{fig2} indicate the times of field 
reversal in each solar hemisphere. It is however difficult to define the exact 
moment of reversal as the timing of reversal depends on the selection of polar area 
\citep{UHa14}.  Both the reversal of magnetic fields, in the latitude range 
45$^{\circ}$\,--\,78$^{\circ}$, and the asymmetry in the reversal between the two 
hemispheres is evident from Figure \ref{fig2}.  It can be seen that the time 
difference in the reversal is of around one year for Cycles 21--23, while it was 
unusually large and around two years for Cycle 24, with the reversal in the North 
having occurred in March 2011, and the reversal in the South having taken place 
in March 2013.  

\begin{figure*}
\centering
\includegraphics[height=6.0cm,width=10cm]{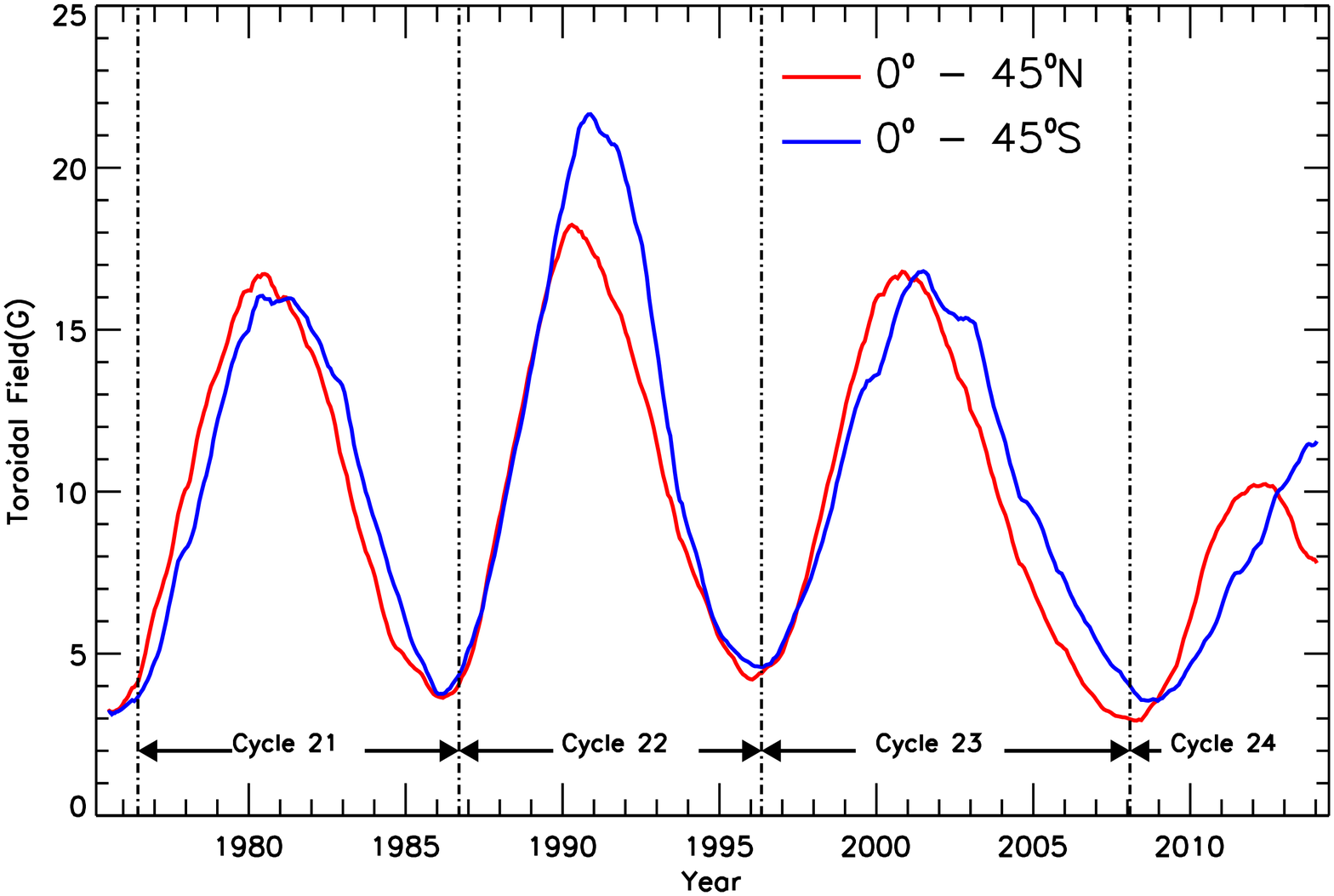}
\caption{The toroidal field in the Northern (red) and Southern (blue) 
hemispheres for solar cycles 21, 22, 23 and 24, obtained using NSO/KP magnetic 
measurements.}
\label{fig3}
\end{figure*}

The hemispheric asymmetry of polar reversal has been primarily attributed to the 
hemispheric asymmetry of solar activity \citep{SKa13} itself.  Figure \ref{fig3} 
shows the variation of toroidal or sunspot fields for solar cycles 21, 22, 
23, and 24 in the latitude range 0$^{\circ}$\,--\,45$^{\circ}$.  These fields 
were estimated from NSO/KP magnetograms as described by \cite{JBG10}.  From 
Figure \ref{fig3}, it is clear that for Cycle 24, the Southern hemisphere reached 
its activity peak $\sim$2 years after the Northern hemisphere.  This hemispheric 
asymmetry in solar activity has presumably led to the observed two year difference 
in the times of field reversals between the two solar hemispheres in Cycle 24.  
In addition, it may be noted that the occurrence of magnetic field reversal 
in both the hemispheres in Cycle 24 (see Fig. \ref{fig1} and Fig. \ref{fig2}) 
indicates that the ongoing solar Cycle 24 is around or past its peak and that 
the declining phase of the cycle has begun.

It is clear from Figure \ref{fig2} that the high latitude fields in Cycle 
24 are comparatively weaker than in Cycles 22 and 23. However, it is the unsigned 
(absolute) high-latitude photospheric field strengths, as shown in Figure \ref{fig4}, 
which showed a remarkable and steady decrease for $\sim$20 years, starting from $\sim$1995. 
The small filled black circles in Fig. \ref{fig4} represent the unsigned 
high latitude fields estimated from NSO/KP magnetograms as described by \cite{JBG10}.  
The large open blue circles in Fig. \ref{fig4} are annual means of the data
with 1$\sigma$ error bars.  The declining trend in the field strength is seen to 
continue after the slight increase during the period 2010\,--\,2011 and the annual means 
for these two years have been shown by large open black circles.  The vertical dotted lines 
in Fig. \ref{fig4} are drawn at solar maximum for each of the 4 cycles 21, 22, 23 and 24.  
An inspection of the behaviour of the annual means in Cycles 21, 22, and 23 indicate that 
the field strength normally declines, as expected, after the solar maximum until the solar 
minimum.  The fact that polar field reversals have taken place in both solar hemispheres in 
Cycle 24 (see Fig. \ref{fig2}) implies that Cycle 24 is now past its maximum and into its 
declining phase.  As a result, the declining trend in the field strength will 
continue at least until 2020, the expected minimum of the current Cycle 24.  The 
solid red line in Fig. \ref{fig4} is a linear fit to the annual means for the period 
1994.48 \,--\, 2014.42, while the dotted red line is a linear extrapolation of the best 
fit line until 2020.  It must be noted that the linear fit used is a least-squares fit to the 
annual means (open blue and black circles in Fig.\ref{fig4}), and not to the actual magnetic 
measurements (black filled circles in Fig. \ref{fig4}).  The solid black line in 
Fig. \ref{fig4} is a linear fit to the annual means after leaving out the two annual means (black 
open circles in Fig. \ref{fig4}) which showed a deviation from the declining trend, in 2010 and 2011.  
The dotted black line is a linear extrapolation to 2020, the expected minimum of the ongoing Cycle 
24.  

The two least square fits shown in Fig. \ref{fig4} are both statistically significant 
with a Pearson' correlation coefficient of r=-0.91, at a significance level of 99\% 
(red solid line) and r = -0.98, at a significance level of 99\%(black solid line).  
Thus, the steady decline which started in $\sim$1995 is expected to go on until 2020, 
(the expected minimum of Cycle 24) that is, for a period of $\sim$ 25 years starting 
from 1995.  From the extrapolation of the two best fit lines, the expected field strength 
in 2020 can be determined.  The field strength will drop to $\sim$1.8 $\pm$ 0.08 G by 
2020 if we consider the fit, in Fig. \ref{fig4}, for the annual means represented by
%
%
\noindent the fitted red line and it will drop to $\sim$1.4 $\pm$ 0.04 G, for the 
annual means represented by the fitted black line.  The two dashed horizontal lines 
in Fig. \ref{fig4} are marked at 1.4 G and 1.8 G and represent respectively, the 
expected field strength in 2020 as derived from the linear extrapolation of the 
black straight line fit, which excludes the two annual means in 2010 and 2011 and the 
red straight line fit, which does not exclude the annual means in 2010 and 2011.

It must be noted that our study of photospheric fields, showing a steady decline 
in the absolute value of the field strength for $\sim$ 20 years, are confined to the 
high latitudes ($\geq$ 45${^{\circ}}$), where sunspots are not present.  However,  
the high-latitude (polar) field and the toroidal field are strongly linked through 
the solar dynamo that causes the waxing and waning of the solar cycle with a period 
of 11 years \citep{Cha10}.  Solar dynamo models assume that the Sun's 
pre-existing poloidal field is transformed to a toroidal field within the Sun, and 
appear as sunspots at the start of the each new cycle.  It is also known that the 
Sun's polar field serves as a seed for future solar activity through their transportation 
by an equatorward subsurface meridional flow \citep{Pet12} from the pole to the equator.  
As a result, polar fields are an important and crucial input in predicting the strength 
of future solar cycles \citep{SPe93,Sch05,SCK05,CCJ07}.  In the following section, we 
have used the high-latitude fields at the cycle minimum to predict the sunspot number 
at the solar maximum of the next cycle.

As mentioned earlier, since the Cycle 24 is already past its peak, the field 
strength is expected to decline until 2020, the minimum of Cycle 24.  Such 
continuously weak high-latitude fields would imply that the toroidal field 
strengths will be also weak and produce weak sunspot fields in subsequent 
cycles.

\section{The Heliospheric Magnetic Field}
   \label{S-HMF}   
The floor level of the HMF is basically the yearly average value to which the HMF 
strength returns to, or approaches, at each solar minimum.  A floor level of the 
HMF exists due to the constant baseline flux from the slow solar wind, and implies 
the presence of magnetized solar wind even in the absence (near-absence) 
of sunspot activity and polar fields.  Since small-scale magnetic fields on the Sun 
are thought to be the source of slow solar wind flows they basically determine the 
floor \citep{CLi11}. In a study of heliospheric magnetic fields (HMF) from 
1872\,--\,2004, \cite{SCl07} had proposed a floor level for the HMF of 4.6~nT.   
In subsequent studies however, measurements of the HMF at 1 AU, have shown a significant 
decline in their strength \citep{SBa08,WRS09,CSS11,CLi11}, going well below 4.6~nT 
during the minimum of Solar Cycle 23.   Figure \ref{fig5} shows (filled black circles) 
measurements of the 27-day averaged values of HMF and (open blue circles) annual 
means of HMF with 1 sigma error bars between 1975\,--\,2014, obtained using OMNI2 
data at 1 AU.  The horizontal dotted line is marked at the proposed floor value 
of the HMF of 4.6 nT \citep{SCl07}, while the monthly averaged sunspot number, scaled 
down (for convenience) by a factor of 10 (SSN/10) from NOAA Geophysical Data Center 
is shown by the solid black curve with the smoothed value overplotted in red.  
The vertical grey bands demarcate 1 year intervals \citep{WRS09} around the minima 
of solar cycles 20, 21, 22 and 23 corresponding to CR1642--1654, CR1771--1783, 
CR1905--1917, and CR2072--2084, respectively.  Average values of the high latitude 
(45$^{\circ}$\,--\,78$^{\circ}$) fields were computed in these 1 year intervals 
for the period 1975\,--\,2014 using NSO/KP synoptic magnetograms. The use of 
these averages will be described subsequently. 

It is evident from Figure \ref{fig5} that the HMF has declined well below the 
proposed floor level of 4.6~nT, and has reached $\sim$3.5~nT during the minimum of Cycle 
23.  Due to the decline of the HMF below 4.6~nT, \cite{CLi11} used the dipole magnetic 
moment derived from polar magnetic field measurements of the Wilcox Solar Observatory 
(WSO) to revise the floor level using two independent methods.  The first was by 
using a correlation between the dipole moment and the HMF at Cycle minimum, while the 
second was by using a correlation between the HMF at each Cycle minimum and the 
yearly sunspot number at Cycle maximum.  These authors reported a revised floor level 
of 2.8~nT \citep{CLi11}.  Using a calibrated database (data from MDI, WSO, and Mount 
Wilson Observatory (MWO) of polar magnetic flux and the HMF from OMNI, \cite{MuS12} 
have also reported the floor level of HMF of 2.77~nT.

We have revisited the floor level in the HMF using high latitude photospheric 
fields and the HMF from OMNI2 database.  It is to be borne in mind that the 
aforementioned studies for finding the floor of HMF are based on the use of 
the signed polar flux (or the dipole moment estimated from the signed polar fields), 
while our studies have used the unsigned high-latitude fields.  The high-latitude 
fields were computed from NSO/KP synoptic magnetograms for the period 1975.14\,--\,2014.42 
and covering Carrington Rotations (CR) CR1625 \,--\, CR2151.  Figure \ref{fig6} shows 
the correlation between the high-latitude or polar field ($B_{p}$) and the HMF ($B_{r}$) 
obtained using values for both $B_{p}$ and $B_{r}$ for 1 year intervals around the 
minima of cycles 20, 21, 22 and 23 (demarcated in Figure \ref{fig5} by grey vertical 
bands).

Based on studies of correlation between polar flux and HMF at solar minimum 
\citep{CLi11,MuS12}, and the fact that the surviving polar fields actually determine 
the floor level of the HMF \citep{WSh13}, we performed a linear fit to find the floor 
of HMF using its correlation with high-latitude fields.  It is to be noted that the 
latter showed a significant decline over the last 20 years.  A linear least square 
fit to the data (Pearson's correlation coefficient of r=0.54 at a significance 
level of 99\% and Spearman's correlation coefficient of r=0.50 at a significance 
level of $\geq$99\%) gave a value of the HMF of $B_{r}$ = 3.2$\pm$0.4 nT, when 
$B_{p}$ = 0 as shown in equation 1.  

\begin{equation}
B_{r}= (3.2 \pm0.4) + (0.43 \pm0.09) \times B_{p}
\label{polarVshmf}
\end{equation}

\begin{figure*}
\centering
\includegraphics[height=6.0cm,width=10.5cm]{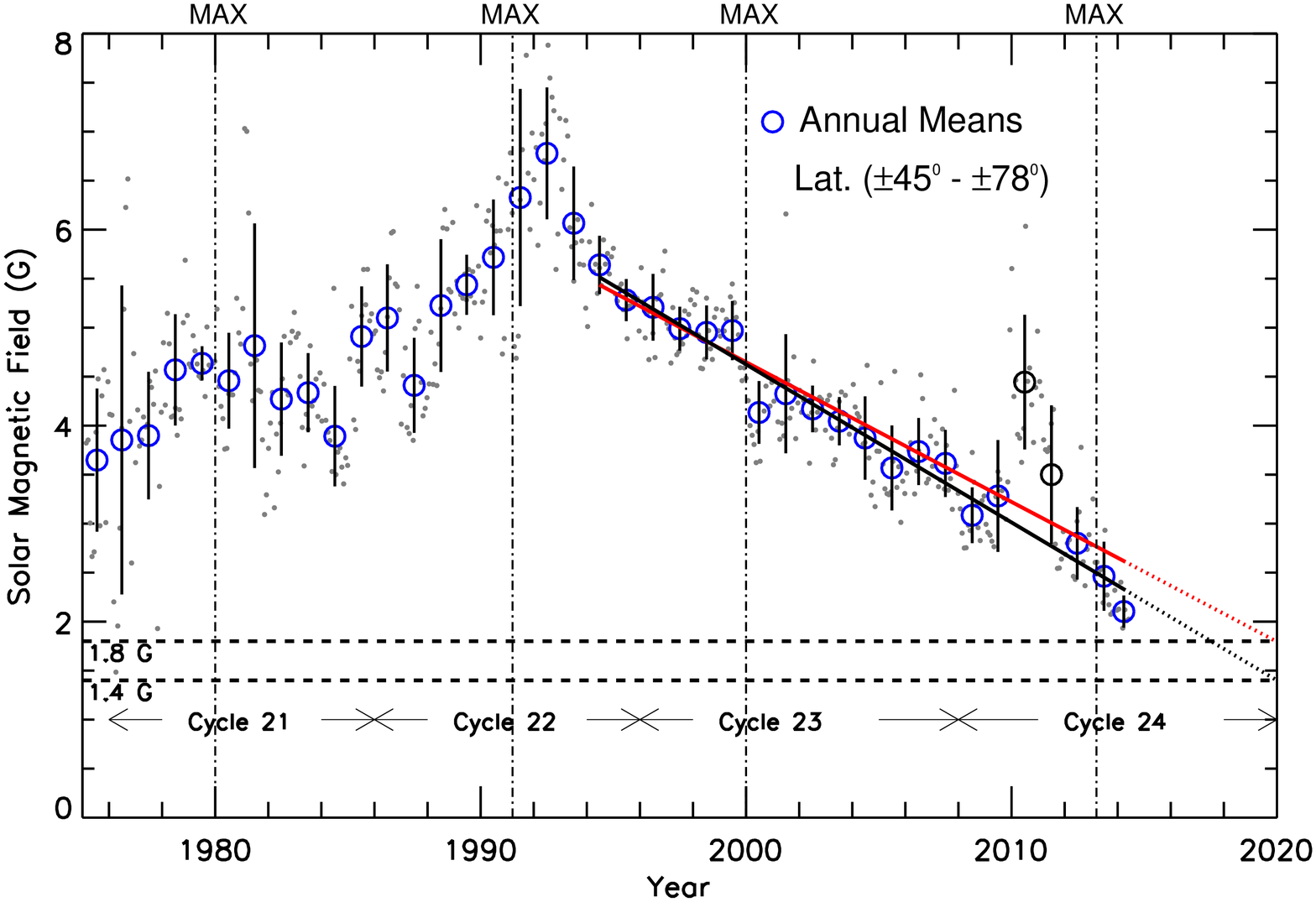}
\caption{The photospheric magnetic fields in the latitude 
range 45$^{\circ}$\,--\,78$^{\circ}$, computed from the NSO/KP 
magnetograms, for the period of 1975.14\,--\,2014.42.  While the solid 
filled dots are actual measurements of magnetic fields, the open 
large blue and black circles are annual means with 1$\sigma$ error bars.  
The solid red line is a best fit to the declining trend for all the annual means, 
while the solid black line is a best fit to the declining trend for the annual 
means after excluding the points of 2010 and 2011 (large black open circles). 
The dotted red and black lines are extrapolations of the two best fit lines 
until 2020, the expected minimum of the ongoing cycle 24.  The horizontal dashed 
lines are marked at 1.8 G the expected field strength at 2020 from the red 
straight line fit and at 1.4 G, the expected field strength by 2020 from the 
black straight line fit. The vertical dotted lines are marked at the respective 
solar maximum of Cycles 21, 
22, 23, and 24.}
\label{fig4}
\end{figure*}

\begin{figure*}[ht]
\centering
\includegraphics[height=6.5cm,width=9.0cm]{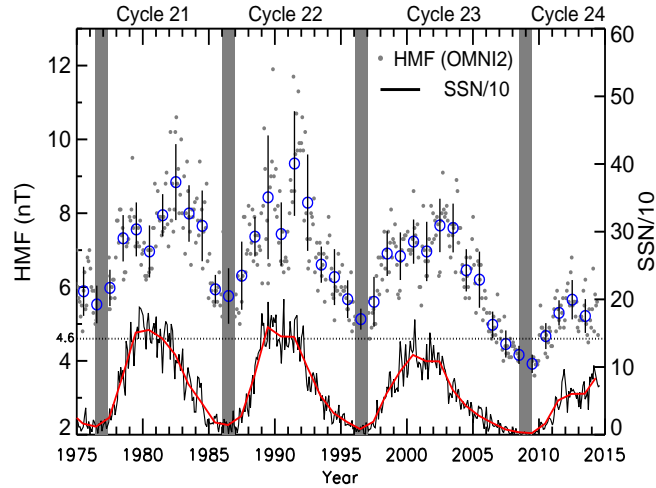}
\caption{Measurements of the heliospheric magnetic field by solid 
black filled dots from OMNI2 data at 1 AU, obtained between 1975\,--\,2014, 
while open circles in blue are annual means with 1 sigma error bars.  
The monthly averaged sunspot number, scaled down by a factor of 10 is 
shown by the solid curve in black with the smoothed value overplotted 
in red. The horizontal line is marked at the floor level of the HMF of
4.6 nT proposed by Svalgaard et al., 2007.  The vertical grey bands 
demarcate 1 year intervals around the minima of solar cycles 20, 21, 
22 and 23.}
\label{fig5}
\end{figure*}

\begin{figure}[ht]
\centering
\includegraphics[height=6.5cm,width=8.0cm]{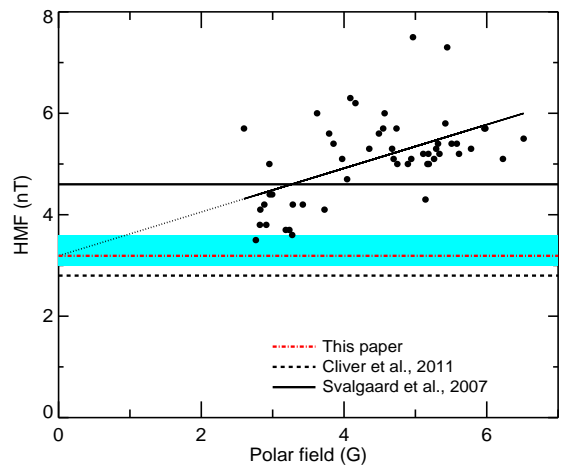}
\caption{A plot of the polar field ($B_{p}$) as a 
function of the HMF for values during 1 year intervals 
around the minima of Cycles 20, 21, 22 and 23.  A linear 
correlation between these two parameters, with a 
correlation coefficient of r=0.54, is shown and gives an 
intercept of 3.2 nT for $B_{r}$ when $B_{p}$ =0.  The 
solid black and dashed black horizontal lines indicate the 
floor levels of the HMF of 4.6~nT, and 2.8 nT respectively, 
as derived by other researchers as described in the text, 
while dotted red line is marked at a floor of 3.2~nT 
derived from the present work. The blue band shows 
the range for our derived floor value of 3.2~nT.}
\label{fig6}
\end{figure}

This implies that even if the polar field, $B_{p}$, drops to zero or to 
very low values, the HMF will persist at a floor level of $\sim$3.2$\pm$0.4~nT. 
The solid black horizontal line in Figure \ref{fig6} is drawn at 4.6~nT, 
the floor level of the HMF proposed by \cite{SCl07}, the dashed black line 
is marked at a floor level of the HMF of 2.8~nT proposed by \cite{CLi11} 
while the dotted red line is marked at a floor of 3.2~nT derived in the 
present work. The blue band shows the range for our derived floor value 
of 3.2~nT. From Figure \ref{fig4} it can be seen that the high latitude 
field drops to between 1.4$\pm$0.04 G to 1.8$\pm$0.08 G in 2020, the 
expected minimum of Cycle 24.  The linear relation obtained between the 
polar field and the HMF (equation 1) implies that the HMF will drop to 
3.9$\pm$0.6 nT by $\sim$2020. The errors of 0.6 was estimated using standard 
formula for propagation of errors.   

A good correlation has been reported (see Fig. 2 of \cite{CLi11}) between 
the peak values of sunspot numbers smoothed over 13-month period ($SSN_{max}$) 
and the HMF at solar cycle minimum given by 

\begin{equation}
SSN_{max}=63.4 \times B_{r} - 184.7 
\label{ssnvshmf}
\end{equation}

Using our estimates of 3.9 nT for the HMF ($B_{r}$) at 2020, the 
expected Cycle 24 minimum in the above linear relationship of $SSN_{max}$ and 
$B_{r}$ of \cite{CLi11}, the $SSN_{max}$ in Cycle 25 can be estimated and 
is likely to be $\sim$ 62 with statistical error bars of $\pm$ 
12 (the estimated standard error of the mean using values of peak sunspot numbers 
for cycles 14--24 from \cite{CLi11}) which is similar with cycle 24 within 
uncertainties.
%
%

\section{Solar Wind Micro-turbulence}
       \label{S-microturbulence}

It is known that photospheric fields during solar minimum conditions generally 
provide most of the heliospheric open flux \citep{SSF00}.  At the solar minimum 
the high-latitude fields extend down to low latitudes into the corona and are 
then carried by the continuous solar wind flow to the interplanetary space to 
form the interplanetary magnetic field (IMF) \citep{SPe93}.  So signatures of any 
change in the long term behaviour of photospheric fields are expected to be 
reflected in the IMF and solar wind.  It has been shown \citep{ACK80} that solar 
wind micro-turbulence levels, as derived from IPS observations of the solar wind, 
are related to both rms electron density fluctuations and large-scale magnetic field
%
%
\protect
%
fluctuations in fast solar wind streams.  There is thus a likelihood of a definite 
causal relationship between the photospheric magnetic fields and micro-turbulence 
in the interplanetary medium thereby implying that a decrease in photospheric 
high-latitude fields will lead to a decrease in micro-turbulence levels in the 
solar wind.

IPS observations of compact extragalactic radio sources provide one with an 
effective and economical ground based method for probing the solar wind plasma 
in the inner-heliosphere both in and out of the ecliptic \citep{ABJ95,JBA97,MoA00}. 
IPS is a phenomenon in which coherent electromagnetic radiation from extragalactic 
radio sources passes through the turbulent, refracting solar wind, considered to be 
confined to a thin screen, and suffers scattering in that process.  This results in
random temporal variations of the signal intensity (scintillation) at the Earth 
\citep{HSW64}, which are modulated by the Fresnel filter function 
$\mathsf{Sin^{2}(\frac{q^{2}\lambda z}{4\pi})}$ where, q is the wave number of the 
irregularities, z is the distance from Earth to the screen, and $\lambda$ is the 
observing wavelength. Due to the action of the Fresnel filter, IPS observations 
are insensitive to larger scale solar wind density fluctuations caused by structures 
such as coronal mass ejections (CMEs).  IPS observations thus enable one to probe 
solar wind electron density fluctuations of scale sizes from 10 to 100 km both in 
and out of the ecliptic \citep{RBA74,CFi85,YaT98,FBD08,ABJ95} and over a wide range 
of distances in the inner-heliosphere \citep{JaB96}.  In fact IPS is sensitive to 
very small changes in the rms electron density fluctuations ($\Delta$N) and has 
even been exploited to study the fine scale structure in cometary ion tails during 
radio source occultations by cometary tail plasma \citep{ARB75,JaA91,JaA92} 
and to study  extremely low density events at 1 AU referred to as solar wind disappearance 
events, when the average solar wind densities at 1 AU dropped to less than 0.1 particles 
cm${^{-3}}$ for extended periods of time \citep{BaJ03, JaF05, JaF08, JDM08}.

The degree to which compact, point-like, extragalactic radio sources exhibit scintillation, 
as observed by ground-based radio telescopes, is quantified by the scintillation index (m) 
given by $\mathsf{m = \frac{{\Delta}S}{<S>}}$, where $\Delta$S is the scintillating 
flux and $<$S$>$ is the mean flux of the radio source being observed. For a given 
IPS observation, m is simply the root mean-square deviation of the signal intensity 
to the mean signal intensity and can be easily determined from the observed intensity 
fluctuations of compact extragalactic radio sources.  Regular IPS observations to 
determine solar wind velocities and scintillation indices at 327 MHz \citep{KKa90,AsK98} 
have been carried out since 1983 from the multi-station IPS observatory, at STEL, Japan.

For a typical IPS observation carried out at a wavelength $\lambda$, the line-of-sight 
to at extragalactic radio source passes through the solar wind at a distance `r' 
defined as the perpendicular distance from the sun to the line-of-sight.  This distance 
r (in AU) is given by Sin($\epsilon$) where, $\epsilon$ is the solar elongation (for a 
schematic of a typical IPS observation, see Figure 1 in \cite{BiJ14b}).  Due 
to the movement of the Earth in its orbit by nearly one degree a day, daily observations 
of a given radio source over an extended period of time will therefore yield observations 
along different lines-of-sight at different distances from the sun thereby enabling 
one to probe the interplanetary medium over a large distance range r at meter wavelengths 
\citep{JAl93, ABJ95,JaB96}.  For an ideal point-like radio source, m steadily increases 
with decreasing r or $\epsilon$ until it reaches a value of unity at some distance of the 
line-of-sight from the Sun.  As r continues to decrease beyond this point, m will again 
drop off to values below unity.  The distance beyond this 'turn-over' can be effectively 
probed by IPS observations.  For sources with extended angular diameters, in the range 
10 to 500 milli arc second (mas), the measured m will be lower than the corresponding 
observation of a point source at a similar solar elongation.  At an observing wavelength 
$\lambda$ of 92 cm or 327 MHz, r ranges from 0.2 to 0.8 AU in the inner-heliosphere and 
IPS is therefore an excellent, cost-effective, technique for studying the large-scale 
structure of the solar wind in the inner-heliosphere.  In the present study, 
to examine the temporal evolution of m for a large number of sources, distributed around 
the sun both in and out of the ecliptic and observed systematically over a long period 
of time, it is necessary to remove both the distance dependence of m and the effect of 
source size to be able to intercompare observations.  In brief, the distance dependence 
of m can be removed by dividing each observation of m, by the corresponding m of a point 
source observed at the same r as described in \citep{JaB11}.  Using many years of systematic 
IPS observations from the Ooty Radio Telescope \citep{SwS71} at 327 MHz, \cite{Man12} have 
established that the strongly scintillating radio source 1148-001 has an angular size 
$\sim$15 milli arc seconds (mas) while VLBI observations have reported an angular diameter 
of 10 mas \citep{VeA85}. Thus, the radio source 1148-001 can be treated as an ideal point 
source and used to remove the distance dependence of m.  A recent study of the temporal 
variations of scintillation index between 1983\,--\,2008, has reported a steady decline 
since $\sim$1995 \citep{JaB11}.

Measurements of the scintillation index, m are basically a measure of the rms 
electron density fluctuations ($\Delta$N), or a proxy for the micro-turbulence 
in the solar wind.  A global reduction in the long-term solar photospheric fields 
\citep{JBG10} would therefore reflect as a corresponding decline in the solar 
wind micro-turbulence levels, as inferred by these IPS measurements. Another recent 
study, covering the Solar Cycle 23, of the solar wind density modulation index, 
$\epsilon_{N} \equiv \Delta{N}/N$, where, $\Delta{N}$ is the rms electron density 
fluctuations in the solar wind and N is the density, reported a decline of around 
8\% \citep{BiJ14b} from 1998 to 2008 which was attributed to the declining 
photospheric fields.  We have therefore re-examined, using IPS observations from 
STEL, the solar wind micro-turbulence levels between 1983\,--\,2013 to see if the 
declining trend has been continuing after 2008.
%
\protect

The effect of the source size can be removed by the method as described in 
\citep{BiJ14b}.  The theoretical values of m for radio sources of a given 
angular size as function of r can be computed using Marian's coefficients 
assuming weak scattering and a power law distribution of density irregularities 
in the IP medium \citep{Mar75}.  For the observed values of m for each given 
source, the best fit Marians curve of a given source size was first determined.  
For example, the best fit Marians curve for 1148-001 corresponds to that obtained 
for a source size of 10 mas.  The observed values of m for each source were then 
normalized by multiplying values of m at each r by a factor equal to the difference 
between the best fit Marians curve for the given source and the best fit Marians 
\begin{figure*}
\centering
\includegraphics[height=6.5cm,width=11.0cm]{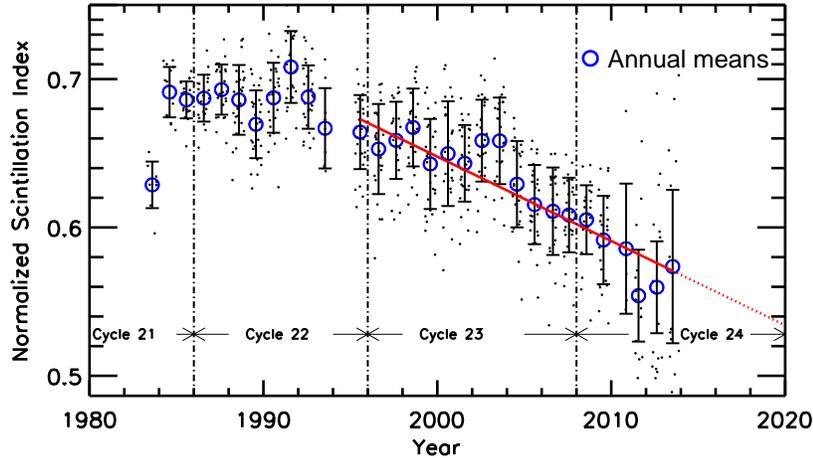}
\caption{m as function of years for observations of 27 
sources after making them both source size and distance independent.  
While the fine black dots show the actual measurements of m, the 
large blue open circles are annual means of m with 1 sigma error bars. The 
solid red line is a best fit to the declining trend while the dotted red line is 
an extrapolation of the fitted line upto 2020, indicated by a 
dashed vertical line.}
\label{fig7}
\end{figure*}
curve for a point source at the corresponding r.  Since 1148-001 is a good 
approximation to an ideal point source, for the present analysis, the observed m 
of all other sources were multiplied by a factor equal to the difference between 
the best fit Marians curve for the given source and the best fit Marians curve for 
1148-001 (see Figure 3 in \citep{BiJ14b} for full description).

After making all the observations of m both distance and source size independent, 
we shortlisted those sources for further analysis which had at least 400 observations 
distributed uniformly (with no significant data gaps) over the entire range of r 
spanning 0.2 to 0.8 AU.  Using the criterion, we have finally shortlisted 26 sources 
from a group of around 200 sources.  The source 1148-001 was included as the 27th 
source though it did not fully comply with the criterion (having few data gaps).  
Figure \ref{fig7} shows, by fine black dots, the temporal variation of m for 
the 27 chosen sources with the large blue open circles representing annual means of 
m with 1 sigma error bars.  It is evident that m continues to drop until the end of 
2013.  The solid red line in Fig. \ref{fig7} is a best linear fit to the 
measurements of m for the period 1995\,--\,2013.  Assuming that this declining trend 
seen in the normalized m continues, we have extrapolated the value of m until 2020, as 
indicated by a vertical red dashed line marked at 2020.  The dotted red line is the 
extrapolation of the best fit line up to 2020.  The steady reduction in the normalized m 
implies that by 2020, a 10 mas `point' source like 1148-001, which is expected to show 
a normalized scintillation level close to unity will show the level of scintillation 
equivalent to that of a 160 mas source implying that that the solar wind 
micro-turbulence level has shown a significant reduction of approximately 30\%.  
\section{Discussion and Conclusion}
 \label{S-Con}
The present study, using synoptic magnetograms from the National Solar Observatory (NSO), 
Kitt-Peak (NSO/KP), USA, has shown a steady decline in photospheric magnetic fields 
at helio-latitudes ${\geq{45^{\circ}}}$, with the observed decline having begun in the 
mid-1990's and continuing, to the end of the dataset through 2014.  In addition, 
heliospheric micro-turbulence levels have also showed a steady decline in sync with 
the solar photospheric fields.  The present, weak solar Cycle 24 coupled with the steady 
and continuing decline in high latitude photospheric fields, starting from $\sim$1995, 
therefore begs the question as to whether we are headed towards a so-called 
``grand minimum" \citep{USK07} similar to the well known Maunder minimum 
(1645\,--\,1715 A.D.) when the sunspot activity was extremely low. There are studies 
that show that peak smoothed annual sunspot number prior to the onset of the Maunder 
minimum was around 50 \citep{Edd76}.  A reconstruction of group sunspot numbers for the 
period 1392\,--\,1985 derived using ${^{10}}$Be ice core records from the North Greenland 
Icecore Project, reported a sunspot number, prior to the onset of the Maunder minimum, 
of around 60 \citep{InK14}. Similarly, a reconstruction of decadal group sunspot numbers using 
${^{14}}$C records from tree rings, shows a cycle averaged group sunspot number, prior to the 
onset of the Maunder minimum, of around 50 at 1610 A.D. \citep{UsH14}. \cite{VaG11} 
showed that the last cycle before the Maunder minimum was small, with sunspot number of about 
20 suggesting for a gradual onset of the Maunder minimum.

Using ${^{14}}$C records from tree rings, sunspot numbers going back over 
the past 1000 solar cycles or $\sim$11000 years in time have been derived and 27 
grand or prolonged solar minima, each lasting on average $\sim$6\,--\,7 solar cycles, 
have been identified in this data set \citep{USK07}.  This implies that conditions 
in those, solar cycles, accounting for about 18\% of the time, were such that they could 
force the Sun into grand minima.  \cite{CKa12} and \cite{KCh13} gleaned very useful 
insights into the process of the onset of grand minima by showing how different 
characteristics of grand minima, seen in their $\sim$11000 year long data set, 
could be reproduced using a flux transport solar dynamo model.  Their study has shown 
that gradual changes in meridional flow velocity lead to a gradual onset of grand minima 
while abrupt changes lead to an abrupt onset.  In addition, 
these authors have reported that one or two solar cycles before the onset of grand minima, the 
cycle period tends to become longer.  It is noteworthy that surface meridional flows 
over Cycle 23 \citep{HRi10} have shown gradual variations from 8.5 m s${^{-1}}$ to 
11.5 m s${^{-1}}$ and 13.0 m s${^{-1}}$ and Cycle 24 started $\sim$1.3 years later than 
expected.  There is also evidence of longer cycles before the start of the Maunder and 
Sp\"{o}rer minimum \citep{MiK10}. Modeling studies of the solar dynamo invoking meridional 
flow variations over a solar cycle have also successfully reproduced the characteristics 
of the unusual minimum of sunspot Cycle 23 and have shown that very deep minima are generally 
associated with weak polar fields \citep{NMM11}. 

A recent analysis of yearly mean sunspot-number data covering the period 1700 to
2012 showed that it is a low-dimensional deterministic chaotic system \citep{ZGk15}.  
Their model for sunspot numbers was able to successfully reconstruct the Maunder Minimum 
period and they were hence able to use it to make future predictions of sunspot numbers.  
Their study predicts that the level of future solar activity will be significantly
decreased leading us to another prolonged or Maunder-like sunspot minimum that will last for 
several decades.  Our study on the other hand, using an entirely different approach, also 
strongly suggests a similar long period of reduced solar activity.    

Our analysis shows that both solar photospheric fields and solar wind 
micro-turbulence levels have been steadily declining from $\sim$1995 and that 
the trend will continue at least until the minimum of Cycle 24 in 2020.  Based 
on the correlation between the high-latitude magnetic field and the HMF at the 
solar minima, we expect that the HMF will decline to a value of $\sim$3.9 
($\pm$0.6) nT by 2020. We also estimate that the peak 13 month 
smoothed sunspot number of Cycle 25 will be $\sim$62 $\pm$ 12, thereby making 
Cycle 25 a slightly weaker cycle than Cycle 24, and only a 
little stronger than the cycle preceding the Maunder Minimum and comparable to cycles 
in the 19th century. It may be noted that, a recent study \citep{ZPo14}, reported that 
the solar activity in Cycle 23 and that in the current Cycle 24 is close to the activity 
on the eve of Dalton and Gleissberg-Gnevyshev minima, and claimed that a Grand Minimum 
may be in progress.  

In an another study however, based on the expected behavior of the axial 
dipole moment after polar reversal in Cycle 24, \cite{UHa14} reported that Cycle 
25 will be similar to Cycle 24.  From our study, the decline in both the 
high-latitude fields and the micro-turbulence levels in the inner-heliosphere since 
1995, which among themselves shows a great deal of similarity in their steadily 
declining trends, again begs the question as to whether we are headed towards 
a Maunder-like grand minimum beyond Cycle 25?

Regarding the floor level of the HMF, it is known that the surviving polar fields 
actually determine the floor level of the HMF \citep{WSh13}.  So if the high-latitude 
photospheric fields continue to decline in the similar manner beyond 2020 and drop 
to very low values, then the HMF will persist at a floor level of $\sim$3.2 nT at 
the expected minimum of the next Cycle 25.  Continued observations through the 
minimum of the current cycle and beyond are therefore crucial to arrive at a decisive 
answer as to whether we are experiencing the onset of a grand minimum.  


%
%
%
%
%
%
%

\begin{acknowledgments}
This work utilizes SOLIS data obtained by the National Solar Observatory (NSO) 
Integrated Synoptic Program (NISP), operated by the Association of Universities 
for Research in Astronomy (AURA), Inc. under an agreement with the NSF, USA. IPS 
observations were carried out under the solar wind program of STEL, Nagoya University, 
Japan.  We thank the NGDC for sunspot data used in this paper. The OMNI data were 
obtained from the GSFC/SPDF OMNIWeb interface at http://omniweb.gsfc.nasa.gov. 
RS and LJ duly acknowledge NASI, Allahabad for support. SA acknowledges an INSA 
senior scientist fellowship.
\end{acknowledgments}


\bibliographystyle{agufull08}
%
%

%
%
%
\end{article}
\end{document}